\documentclass{aa}  

\usepackage{natbib}
\bibpunct{(}{)}{;}{a}{}{,} 
\usepackage{xcolor}
\usepackage[normalem]{ulem}
\usepackage{graphicx}

\usepackage{lipsum}
\usepackage{lscape}             
\usepackage{placeins}           

\usepackage{txfonts}

\usepackage[colorlinks=true,allcolors=blue]{hyperref}

\usepackage{subfigure}
\usepackage{subcaption}

\begin{document}

   \title{Bent-tailed radio galaxies as tracers of the intrafilament medium}

   \author{Katlego Morakile\inst{1}
          \and Kshitij Thorat\inst{1} 
          \and Etienne Bonnassieux\inst{2,3} 
          \and Tiziana Venturi\inst{3,4}
          \and Konstantinos Kolokythas\inst{4,5,3}      
          }
          
  \institute{Department of Physics, University of Pretoria, Private Bag X20, Hatfield 0028, South Africa \\
  \email{katlegomogamis@gmail.com}
         \and
             Laboratoire d’Astrophysique de Bordeaux, Univ. Bordeaux, CNRS, UMR 5804, 33615 Pessac, France
        \and
          INAF – Istituto di Radioastronomia, via Gobetti 101, I–40129 Bologna, Italy 
        \and
           Centre for Radio Astronomy Techniques and Technologies, Department of Physics and Electronics, Rhodes University, P.O. Box 94, Makhanda 6140, South Africa  
        \and
            South African Radio Astronomy Observatory, 2 Fir Street, Observatory 7925, South Africa}

   \date{Received Month DD, YYYY; accepted Month DD, YYYY}

  \abstract
   {Bent-tailed radio galaxies have traditionally been associated with, and used as tracers of, the dense intracluster medium. In recent years, however, such sources have also been identified outside of clusters, with only a few discovered in low-density environments of cosmic filaments. Their occurrence in low-density environments remains rare and is still poorly understood.} {We announce the discovery of bent-tailed radio galaxies in galaxy cluster filaments. We have identified two bent-tailed radio galaxies, J0020-25 and J0019-26, within a filament at the redshift of $z \sim 0.1273$, and another bent-tailed radio galaxy, J0019-25, in a separate filament at $z \sim 0.1126$. We used MeerKAT observations of these bent-tailed radio galaxies to place constraints on their intrafilament medium (IFM).}{ The effects of the filaments could be seen in the radio morphologies of these galaxies. Using faint regions in each galaxy, we estimate the intrafilament pressure, density, magnetic field strength, and total energy density.}{We found an IFM density of $\rho_\text{IFM} \approx 3.8\times 10^{-29}$ g cm$^{-3}$ for the surrounding medium of J0020-25 and J0019-26, and $\rho_\text{IFM} \approx (0.2 - 1.1) \times 10^{-26}$ g cm$^{-3}$ around J0019-25. We found that our estimates for the intrafilament environment around J0020-25 and J0019-26 lie within the $\rho_\text{IFM}$ literature estimates, while the density estimate found using J0019-25 is higher than the literature density estimates for the IFM and in agreement with the density of group environments. J0019-25 likely resides in a galaxy group embedded within a filament. These discoveries show the capabilities of using bent-tailed radio galaxies as probes to locate the otherwise invisible large-scale filaments.}{}

   \keywords{radio galaxies -- active galactic nuclei -- galaxy filaments -- intrafilament medium}

   \maketitle

\section{Introduction}

   Jetted radio galaxies are active galaxies that host an active galactic nucleus (AGN) at the centre and emit collinear radio jets on opposite sides of the AGN. However, some radio galaxies show deviations from the alignment of the two opposing jets, to form tails of radio emission. Bent-tailed radio galaxies (BTRGs) are jetted radio galaxies with non-collinear radio jets. They are classified on the basis of the angle between the two bent jets. A bent-tailed radio galaxy (BTRG) with an angle greater than 90\textdegree\ between the two bent radio jets is called a Wide Angle Tailed (WAT), whereas the one with an angle less than 90\textdegree\ between the two bent radio jets is called a Narrow Angle Tailed (NAT) (\citealp{owen1975radio, odea1985global}; for a recent review on WATs, see \citealp{odea2023wide}). The last group includes the head-tail (HT) radio galaxies \citep{owen1975radio}. HT radio galaxies are characterised by a bright radio 'head' associated with the host galaxy and a faint tail of radio emission trailing behind them \citep{owen1975radio}. These deviations of radio galaxy morphologies from the expected linear structure are believed to be due to galaxy interaction with the environment. Understanding the physical mechanisms behind bending in BTRGs remains an active area of research, with current constraints largely attributed to insufficient angular resolution to trace jet bending close to the core.
   
   To date, several mechanisms have been attributed to jet bending. These include: ram pressure, caused by the relative movement of the galaxies through the medium \citep{gunn1972infall, Garon2019, Bhukta2021, morris2022does, Vardoulaki2025} or buoyancy, caused by deviations in the density of the intragalactic medium, which interacts with the intracluster medium \citep{cowie1975dynamical, odea2023wide}. Other environmental factors that cause radio jet bending include turbulence in the environment, shocks, stellar winds, and interactions of radio jets with neighbouring galaxies \citep{Garon2019, Parekh2020FRI} that can give rise to sloshing motions (e.g., \citealp{Kolokythas2020}). 

    Several studies have associated BTRGs with galaxy clusters, suggesting that they can be used as tracers of galaxy clusters (e.g., \citealt{Blanton2000, smolvcic2007wide, Dehghan2014}). However, there have been studies that have identified BTRGs outside galaxy clusters, in galaxy groups and filaments (e.g., \citealt{freeland2008bent, edwards2010first,  OBrien2018, Garon2019, Vardoulaki2021}). The observed trend from these studies is that BTRGs are associated with overdensities, and not with galaxy clusters specifically.  

    The presence of BTRGs within filamentary structures serves as a promising tool in current astrophysical research. Galaxy filaments are the largest structures in the cosmic web, acting as bridges that connect galaxy clusters. They contain gas, galaxies, and dark matter, separated by vast, empty voids. They have lower densities and temperatures compared to galaxy clusters. The Warm-Hot Intergalactic Medium (WHIM), which contains about half of the universe's baryonic matter, resides in these filaments. The WHIM has cooler temperatures of 10$^5$-10$^7$ K and low density \citep{Cen&Ostriker, malarecki2013giant}. Due to its warm-hot temperature, it cannot be detected with existing instruments. Its temperature is not hot enough to be detected in X-ray and is also not cold enough to be detected in H\textsc{I}. 
    
    Previous claims of filamentary WHIM detections relied on indirect methods, such as the stacking technique and looking at the spatial distribution of galaxies. These include \citet{deGraaff2019}, who stacked the thermal Sunyaev-Zel’dovich (tSZ) effect for pairs of galaxies in the Sloan Digital Sky Survey; \citet{Tanimura2020FirstDetection, Tanimura2022} who made the first and follow-up detections of stacked X-ray emission from filaments; \citet{Vernstrom2021} who used stacked radio and X-ray emission, and further confirmed the stacked X-ray filament signals from \citet{Tanimura2020FirstDetection}; and \citet{Li2026JCAP} using tSZ and CMB lensing stacking. These studies have revealed filamentary signals at multiple wavelengths, providing statistical evidence of the otherwise faint WHIM gas. Other studies followed the spectroscopic approach, detecting the missing baryons through the identification of OVII and OVIII absorbers on the quasar spectra via X-ray observations \citep{Nicastro2018, bonamente2025}, as well as the forbidden line of [Fe XXI]\ $\lambda \ 1354 \ \AA $ at UV wavelengths \citep{Fresco2020}.  

    Recent statistical analyses have shown that filaments play a primary role in directing the flow of matter and shaping the properties of the galaxies and groups they host. For instance, \citet{Tempel2014} demonstrated that galaxies and groups are distributed along filaments in regular patterns, resembling a "pearl necklace", where the "pearls" correspond to galaxy groups. This suggests that the occurrence and positions of galaxies, including bent-tailed radio galaxies, may not be random but related to their locations with respect to groups and overdensities within this regular filamentary pattern. This suggests that the large-scale gravitational environment influences galaxy evolution on a fundamental level. Furthermore, \citet{Bonjean2018}, in their study of the A399-A401 inter-cluster filament, found that galaxies in the filament have similar star formation rates to those already inside the clusters. The authors found a similar population of passive, red, and dead galaxies in both environments. This suggests that the star formation quenching process can begin while a galaxy is still travelling through a filament, indicating that these environments are active enough to impact galaxies before they ever reach a cluster. 
    
    This early transformation in galaxies is often referred to as pre-processing \citep{Fujita2004}. As discussed by \citet{Sarron2019}, filaments act as primary sites where galaxies are processed before entering larger cluster environments. From the perspective of the morphological peculiarities of jetted radio galaxies, the bending of jets in a filamentary environment is attributed to the ram pressure exerted on the galaxies by the intrafilamentary medium (IFM) as the host galaxies move through it. Within filaments, WHIM low density usually results in very slight, large-scale jet curvature. However, very high velocities of the host galaxies can compensate for these low densities, generating enough ram pressure to bend the jets. The bending effects often become more prominent as a galaxy transitions from a filament into a galaxy group, encountering the denser intragroup medium. These effects on a radio jet can sometimes mark the exact boundary of a galaxy group, serving as proxies for identifying the virial radius of a galaxy group within a filament. Consequently, the presence of BTRGs within filamentary structures can serve as probes to identify and characterise the physical properties of the otherwise invisible WHIM. 

    Depending on the orientation and location of the galaxy with respect to the filament, the jet-IFM interaction can result in bent, complex, or asymmetric morphologies. This has been shown in a study by \citet{Mohato2026}, who performed a statistical study to check if there is any correlation between the filamentary spine and the orientation, proximity, and size of giant radio galaxies (GRGs) and standard radio galaxies. The authors found that galaxies with complex morphologies -- including those that are hybrid Fanaroff--Riley (FRI/II; \citealp{fanaroff1974morphology}) or too ambiguous in morphology -- tend to be found near the filament spine, while WATs are located farther from the filament spine, suggesting that the IFM dynamics affect the morphology and properties of the galaxies.

    Studies such as \citet{freeland2008bent} and \citet{edwards2010first} have demonstrated that BTRGs can be used to estimate the properties of their ambient medium. The cocoons of active galaxies (low flux density regions at the edges of the radio lobes) have been proven to have very low energies, allowing us to assume pressure equilibrium with the environment \citep{Cotton2020}. Relaxed components of FRII radio galaxies, also assumed to be in pressure equilibrium with the surrounding ambient medium, have been used to estimate the properties of that medium \citep{malarecki2013giant, Peng2015GRGs, Saripalli2015, Oei2023}.
    
    In this work, we performed a follow-up of a study by \citet{Pimbblet2005}, who used the spatial distribution approach and found a `wall' over $>$ 40 $h^{-1}$ Mpc long in the redshift range of 0.1126 $< z <$ 0.1188 (33750 $< cz <$ 35625) towards Abell 22. Their findings suggest that this wall is part of a 10 $h^{-1}$ Mpc wide large-scale filament running between two Abell clusters, namely Abell 47 and Abell 15, on either side of Abell 22. Abell 22 is a galaxy cluster at $z = 0.142$. We used a MeerKAT image of the Abell 22 galaxy cluster, from which we identified a BTRG associated with the wall identified by \citet{Pimbblet2005} and two other BTRGs associated with other filamentary structures towards Abell 22. We used the BTRGs found in all the identified filamentary structures towards Abell 22 to estimate the properties of the intrafilament medium. This study further demonstrates that BTRGs can be found in cosmic filaments and can act as probes of their environment. Throughout this work, we assumed a flat Lambda Cold Dark Matter ($\Lambda$CDM) cosmology with Planck18 cosmological parameters $H_{0} = 67.7$ km s$^{-1}$ Mpc$^{-1}$, $\Omega_{m} = 0.31$, and $\Omega_{\Lambda} = 0.7$ \citep{Planck2020}.

\section{Data products}
\label{sec:Data Products}

   An archival Abell 22 cluster image from the MeerKAT Galaxy Legacy Survey (MGCLS) was used for this study. MGCLS is a survey of 115 galaxy clusters observed using the MeerKAT telescope. The survey was observed using MeerKAT’s L-band receiver in the 4K correlator mode. The observations were made in full polarisation between -80° and 0° declination and a full range of right ascension, with $\sim$6-10 hours of observation time per cluster \citep{Knowles2022}.  

    MGCLS\footnote{\url{http://mgcls.sarao.ac.za/data-releases/}} data products release (DR1) has been made public and consists of two different products, the basic and enhanced data products. The basic products consist of image cubes with 8" resolution and no primary beam corrections. Enhanced products consist of the five plane primary-beam-corrected image cubes at a full resolution of $\sim$8" and a convolved resolution of 15". The five plane image cubes consist of the brightness at the reference frequency, spectral index, the brightness uncertainty, the spectral index uncertainty, and the $\chi^2$ of the least squares, respectively. 

    The basic products span 2\textdegree\ $\times$ 2\textdegree\ full field of view, while the enhanced products span only the inner 1.2\textdegree\ $\times$ 1.2\textdegree\ of the field of view. The MGCLS products have an average image noise of $\sim$3-5 $\mu$Jy at full resolution. These products also come with non-blanked primary-beam-corrected frequency cubes, which are 12 plane frequency cubes with centre frequencies 908, 952, 996, 1044, 1093, 1145, 1318, 1382, 1448, 1482, 1594, and 1656 MHz. More information on the MGCLS observations, calibration, catalogues, and science merits can be found in \citet{Knowles2022}.  

    We used optical data from the Las Campanas/ Anglo-Australian Telescope Rich Cluster Survey (LARCS) spectroscopic cluster catalogue, Two-degree Field (2dF) Galaxy Redshift Survey (2dFGRS), and the Dark Energy Spectroscopic Instrument (DESI) Legacy Imaging Surveys to obtain the redshift information for the galaxies in the region of interest.

\section{Galaxy distribution}

\begin{figure}
    \centering
    \includegraphics[width=0.82\linewidth]{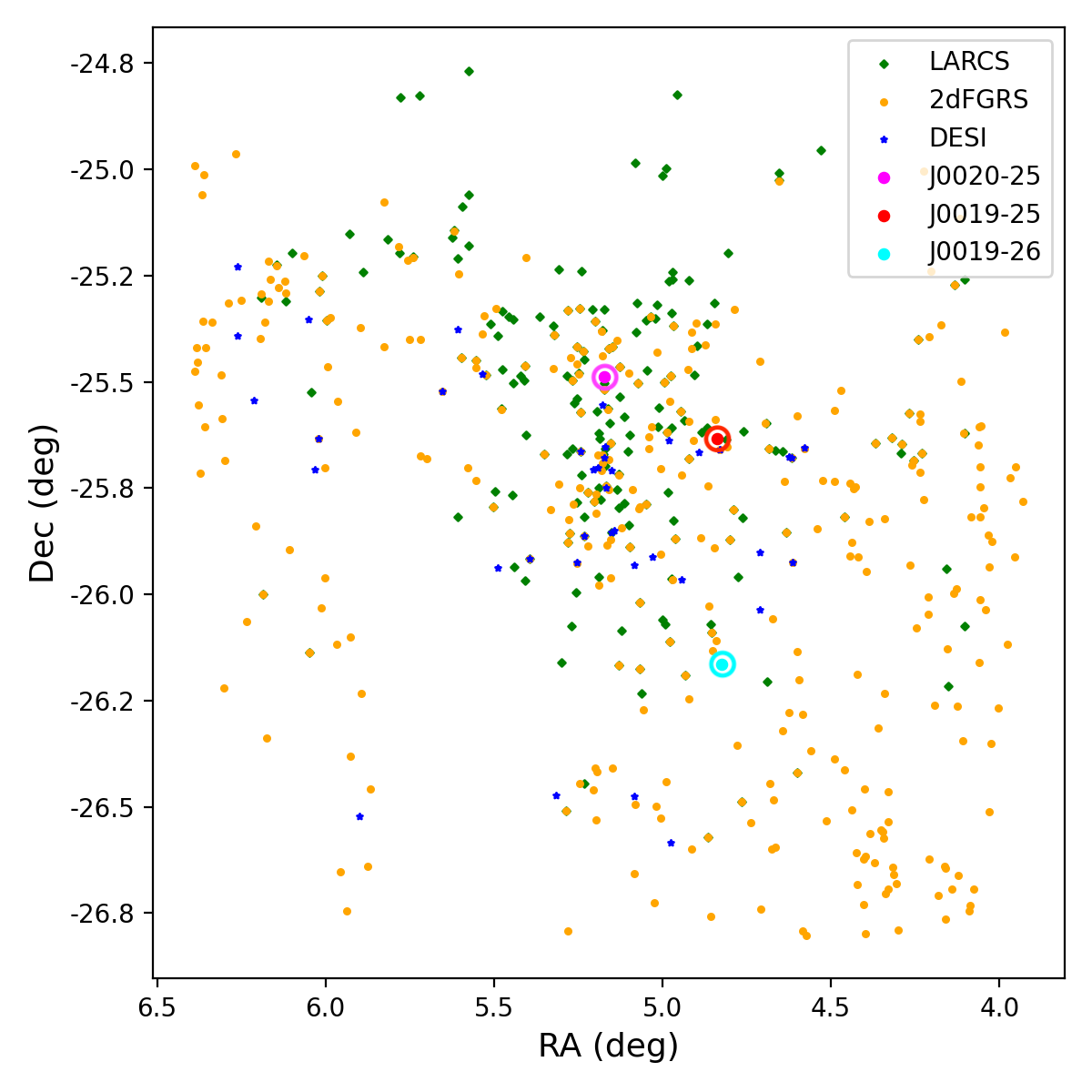}
    \caption{Spatial distribution of galaxies in the direction of Abell 22. In green, we have LARCS optical sources; in orange, we have 2dFGRS sources; and in blue, we have DESI sources. We also show the positions of J0020-25 (red), J0019-25 (red), and J0019-26 (cyan) on the plot.}
    \label{fig: Spatial distribution}
\end{figure}

To follow up on the `wall' identified by \citet{Pimbblet2005}, we used optical data to look at the distribution of galaxies in the direction of the identified large-scale filament. We looked at the distribution of galaxies with up to the redshift of $z = 0.15$, and within the RA and Dec range of 00:15:42.9 $\leq$ RA $\leq$ 00:25:34.3, and -24:37:02.1 $\leq$ Dec $\leq$ -26:48:59.7. Figure \ref{fig: Spatial distribution} shows the spatial distribution of galaxies within our specified limits.

\begin{figure}
    \centering
    \includegraphics[width=0.82\linewidth]{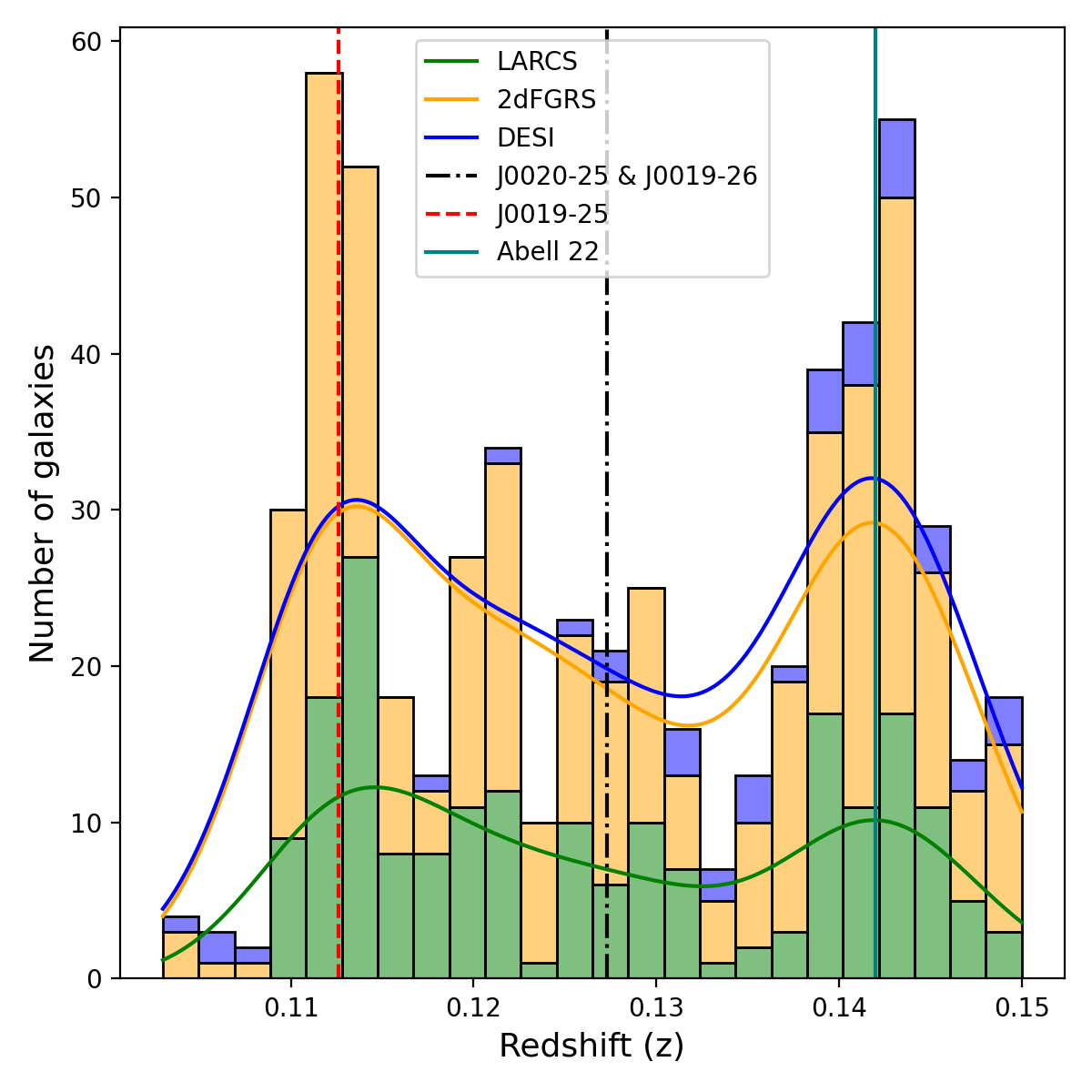}
    \caption{Redshift distribution of galaxies in the direction of Abell 22 with bin width of 0.002, within the redshift range of $z \in [0.10, 0.15]$. The black vertical line shows the position of J0020-25 and J0019-26, while the red vertical line shows the position of J0019-25 within the distribution. The solid continuum lines represent the Kernel Density Estimation (KDE) of each sample, providing a smoothed estimate of their probability density functions.}
    \label{fig: Redshift distribution}
\end{figure}

\begin{figure}
    \centering
    \includegraphics[width=0.82\linewidth]{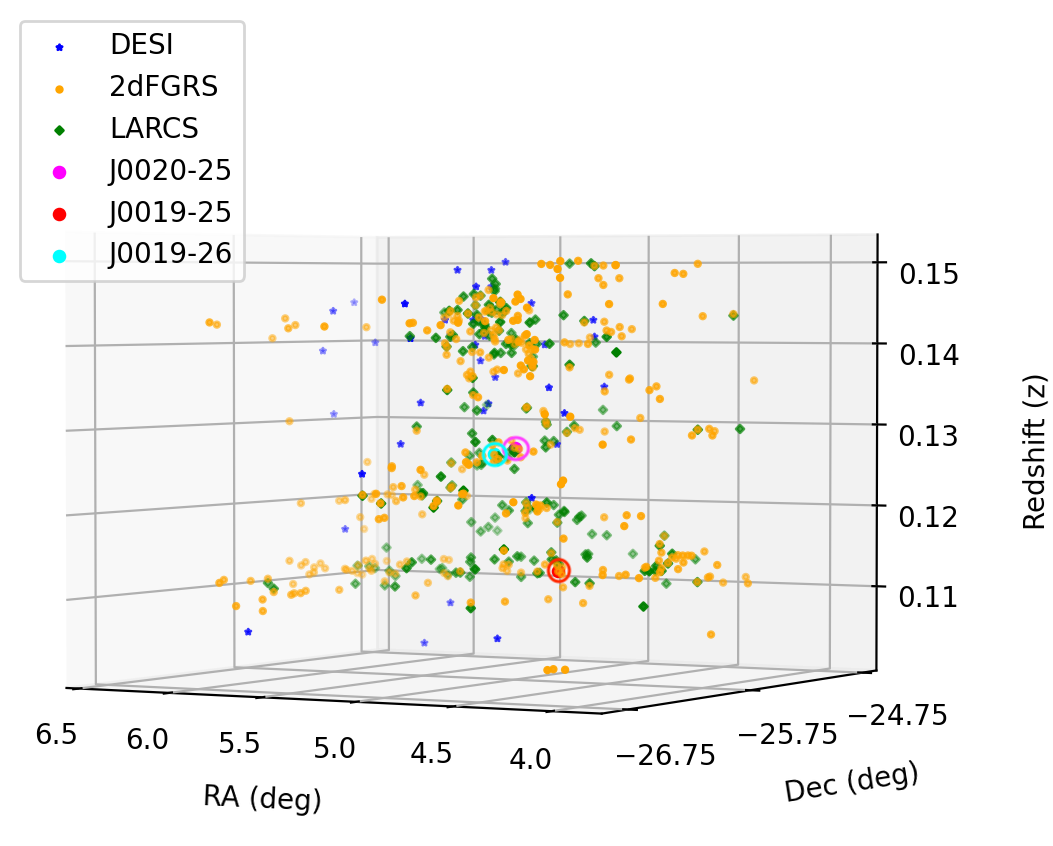}
    \caption{3D distribution of galaxies towards Abell 22. The red, red, and cyan dots show the locations of J0020-25, J0019-25, and J0019-26 in the 3D space, while the dense circular region at $z = 0.142$ represents Abell 22.}
    \label{fig:3D distribution}
\end{figure}

Within the field of view of the MeerKAT image used in this study, we identified three bent-tailed radio galaxies, i.e. J0020-25, J0019-25, and J0019-26. The positions of these galaxies are shown in all of the distribution plots (Fig. \ref{fig: Spatial distribution}, \ref{fig: Redshift distribution}, and \ref{fig:3D distribution}). Figure \ref{fig: Redshift distribution} shows the histogram used to examine the redshift distribution, providing insights into the universe's large-scale structure and allowing us to identify dense environments.  

The redshift distribution on the histogram shows a high concentration of galaxies around $z \in [0.111, 0.115]$ and $z \in [0.138, 0.144]$. The elevated number of galaxies around $z \in [0.111, 0.115]$ supports the findings of \citet{Pimbblet2005} that there exists a wall at $z \in [0.112, 0.118]$. The concentration of galaxies around $z \in [0.138, 0.144]$ coincides with the Abell 22 redshift of $z = 0.142$. To further investigate this, a 3D plot was used to show the 3D distribution of the galaxies towards Abell 22 (Fig. \ref{fig:3D distribution}). It shows a well-defined large-scale structure with a prominent horizontal band extending over the redshift range $z \sim 0.110$ to 0.114 and a more vertically inclined band that spans approximately $z \sim 0.120$ to 0.130. These two structures are connected by a bridge of galaxies, suggesting the presence of multiple intersecting filaments in the direction of Abell 22. The galaxies J0020-25, J0019-25, and J0019-26 are all located along these filamentary features. 

\section{Sample}
\subsection{J0020-25}
\begin{figure} 
    \centering
    \begin{subfigure} 
        \centering
         \includegraphics[width=\linewidth]{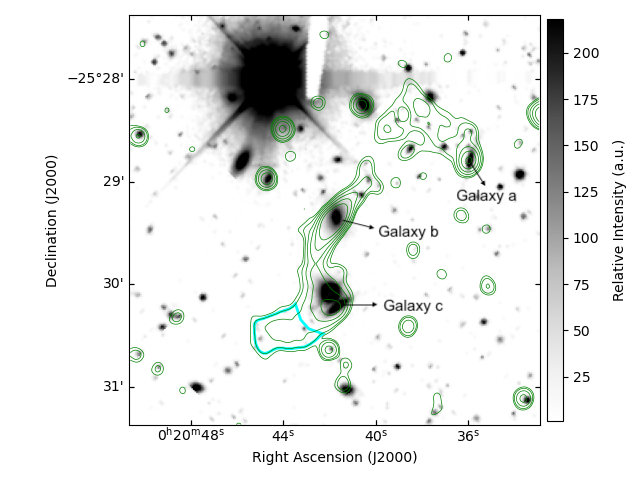}\vspace{-0.6cm}
         \caption*{J0020-25} 
         \label{fig:J0020-25}
    \end{subfigure}
    \hfill
    \begin{subfigure}
        \centering
        \includegraphics[width=\linewidth]{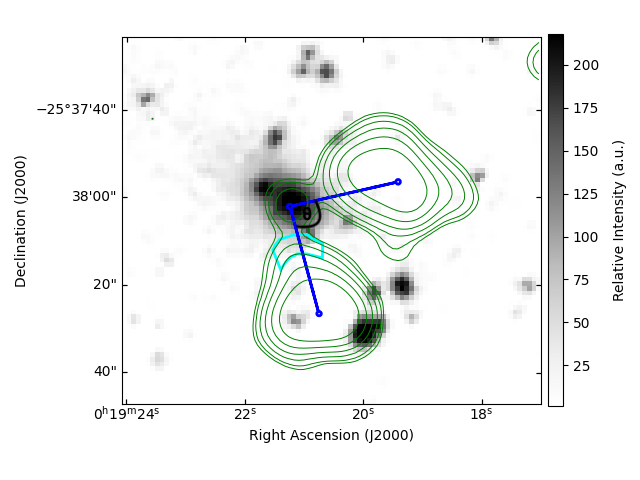}\vspace{-0.8cm}
        \caption*{J0019-25} 
        \label{fig:SecondGalaxy}
    \end{subfigure}
    \hfill
    \begin{subfigure} 
        \centering
        \includegraphics[width=\linewidth]{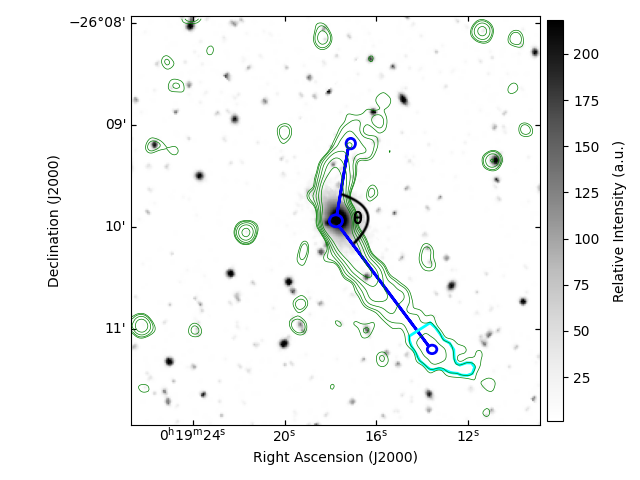}\vspace{-0.6cm}
        \caption*{\label{fig:ThirdGalaxy}J0019-26}
        
    \end{subfigure}

    \caption{J0020-25 DES-DR2 (top), J0019-25 PanSTARR (middle), and J0019-26 DES-DR2 (bottom) optical relative intensity images with $6.6\ \mu$Jy/beam $\times$ 3, 5, 10, 20, 40, and 80 radio contours overlaid (in green). Cyan outlines faint regions used for IFM property estimation. We also illustrate the lines used to estimate the bending angle (blue), where theta ($\theta$) is the bending angle of the galaxy.}
    \label{fig:AllGalaxies}
\end{figure}

J0020-25 has an unusual morphology. It looks like a bent-tailed with a diffuse, possibly disrupted northern lobe (Fig. \ref{fig:AllGalaxies} - top panel), which may indicate a denser environment in the northern direction of the source. According to its radio brightness distribution, J0020-25 seems to have a core at (RA, Dec) = (0:20:41.8, -25:29:22.6) with an optical counterpart of redshift $z = 0.1273$, corresponding to 2dFGRS TGS138Z182 with a B band magnitude of 18.059  \citep{colless20012df}. It has an estimated luminosity of L$_{0.150GHz}$ = 4.17$\times 10^{24}$ W\,Hz$^{-1}$, suggesting that it is an FRI-type galaxy, as according to the revised FRI/FRII luminosity break of L$_{0.150GHz}$ = $10^{26}$ W\,Hz$^{-1}$ by \citet{mingo2019revisiting}. Looking at the morphology of J0020-25 radio emission, the following questions arise: Is the northern emission part of this galaxy? If so, what might have disrupted this lobe? If not, where is the north lobe of the observed galaxy? 

We looked into the radio-optical cross-match of this galaxy (Fig. \ref{fig:AllGalaxies} - top panel) from which we identified multiple optical sources (Galaxy a, b, and c, where Galaxy b is the host galaxy of J0020-25) within its radio emission. Through the SIMBAD database, the optical galaxy in the disrupted north lobe (Galaxy a) was found to have a redshift of $z = 0.347$, showing that it is a background galaxy. The redshift of Galaxy c ($z = 0.140$) suggests that it is also a background galaxy of J0020-25 and possibly part of the Abell 22 galaxy cluster. Cross-matching the "south lobe" emission with optical data, we find that the peak of the south lobe emission does not correspond with Galaxy c. This offset suggests that the observed emission is not physically associated with Galaxy c, but represents the south lobe emission from J0020-25. Although redshift measurements are crucial for a conclusive statement, the morphology alignment and position offset favour an association with J0020-25.

\begin{figure*}
  \centering
  \begin{minipage}[b]{0.48\textwidth}
    \centering
    \includegraphics[width=\linewidth]{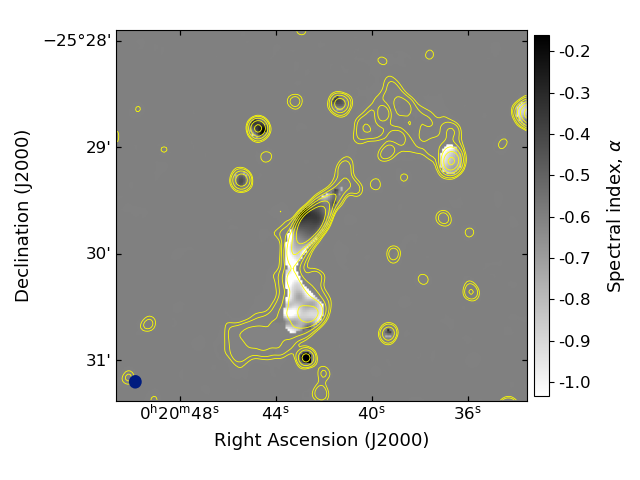}
    \centerline{(a) J0020-25.}
  \end{minipage}
  \hfill
  \begin{minipage}[b]{0.49\textwidth}
    \centering
    \includegraphics[width=\linewidth]{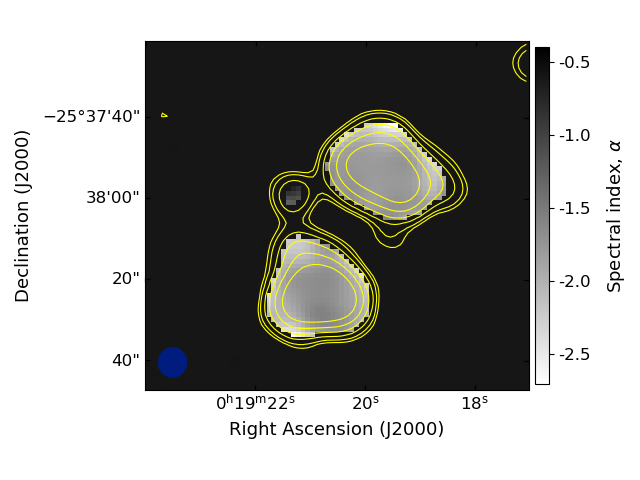}
    \centerline{(b) J0019-25.}
  \end{minipage}
  \vspace{6pt}
  \caption{Spectral index maps extracted from the second plane of the MGCLS full resolution (7.62" $\times$ 7.22") Abell 22 enhanced product, overlaid with full resolution radio emission contours (yellow) at $6.6\ \mu$Jy/beam $\times$ 3, 5, 10, 20, 40, and 80 levels. The colour bars indicate the spectral index values ($\alpha$), with the background fixed to the default MGCLS value of -0.6.  The synthesised beam is indicated by the ellipse in the bottom left corner of each map.}
  \label{fig:Spectral index maps}
\end{figure*}

We also looked at the spectral index map of this source from the second plane of the MGCLS Abell 22 enhanced product (Fig. \ref{fig:Spectral index maps}). The map shows a flat spectrum at the core, which typically indicates a younger or freshly injected population of electrons. We also see a steep spectrum towards the south lobe, which may indicate older electron populations that have undergone energy losses. While this map traces the spectral trends in the brighter regions, the low surface brightness regions in the tails suffer from low signal-to-noise ratios, resulting in poor spectral tracing.

\subsection{J0019-25}
The second galaxy, J0019-25, is found at (RA, Dec) = (0:19:21.2, -25:38:01.6). It has an optical host 2MASX J00192126-2538012, with a redshift of $z = 0.1126$ \citep{colless20012df} and a K magnitude of 13.163 \citep{Skrutskie2006} found through SIMBAD. It has an estimated luminosity of L$_{0.150GHz}$ = 2.25$\times 10^{25}$ W\,Hz$^{-1}$, suggesting that it is an FRI-type radio galaxy, as according to \citet{mingo2019revisiting}. This galaxy could also be classified as a wide angle tailed galaxy. It has a steep and ultra steep spectrum at the core and the lobes, respectively (Fig. \ref{fig:Spectral index maps}). The steep spectral index at the core may indicate that the AGN has recently switched off and stopped injecting fresh relativistic particles into the jets. The absence of the freshly injected plasma is also seen through the ultra-steep spectral index in the radio lobes. Higher angular resolution would be needed to resolve the lobes and test for the presence or absence of the hotspots or high brightness peaks, to confirm the nature of this galaxy.

\subsection{J0019-26}
The third galaxy, J0019-26, is found at (RA, Dec) = (0:19:17.8, -26:09:56.8). Its optical host is 2dFGRS TGS137Z256 with the redshift of $z = 0.1273$ \citep{mauch2007radio} and K band magnitude of 12.633 \citep{Skrutskie2006}. The morphology of this galaxy (Fig. \ref{fig:AllGalaxies} - bottom panel) suggests that it is a wide angle tailed FRI radio galaxy. It has a luminosity of L$_{0.150GHz}$ = 8.19$\times 10^{24}$ W\,Hz$^{-1}$, which suggests that it is an FRI radio galaxy, supporting the observed morphology \citep{fanaroff1974morphology}.

\section{Methods}
\subsection{Bending angle}\label{sec:Bending Angle}
In this work, we define the bending angle ($\theta$) of a galaxy as the angle between the two bent radio jets. We estimated the bending angle of J0019-25 and J0019-26 using the peak ﬂux density method (\citealp[also used in][]{Garon2019, Vardoulaki2025}), measuring the position angles of the vectors joining the host positions and the peak flux density in the radio lobes. The peak flux density positions within the radio lobes typically represent the hotspots, which allow us to trace the most recent jet axis. This approach was selected due to its objectivity in identifying the jet axis and has been shown to be statistically robust and reliable by \citet{Vardoulaki2025}. While this method is highly reliable for well-defined sources, it can be difficult to apply to objects that lack a clear peak flux density position. For J0019-25 and J0019-26, the bending angle was taken to be the difference between the position angle of the two vectors representing the northern and southern jets. The lines used to determine $\theta$ are shown in Figure \ref{fig:AllGalaxies}, with the estimated bending angles summarised in Table \ref{table:galaxy_properties}. 

\subsection{Spectral index analysis}
\label{subsec:spectral analysis}
Our attempts to produce full 2D spectral index maps using Broadband Radio Astronomy Tools (\textsc{brats}; \citealp{Harwood2013}), and the full resolution 2D maps provided with the MGCLS products, yielded poor, non-detailed maps for the radio tails, driven by the low surface brightness and limited signal-to-noise ratios in these extended regions. Additionally, J0019-26 lacks a spectral index map from the MGCLS enhanced products. Due to poor spatially resolved spectral index maps and to ensure a uniform and reliable analysis across our sample, we estimated integrated spectral indices for our targets directly from the frequency cubes provided in the MGCLS enhanced data products \citep{Knowles2022}. We derived the integrated flux density spectral indices ($\alpha$) by performing an independent power-law fit ($S_{\nu} \propto \nu^{\alpha}$) to the flux densities measured across the 12 plane frequency cubes, as described in Section \ref{sec:Data Products}.

\subsection{Intrafilament pressure}
\label{subsec:pressure_methods}
Once the AGN stops feeding the lobes with fresh, high-energy electrons, there will be no generation of a new population of electrons. The existing electron population will lose energy with time until it reaches pressure equilibrium with the environment, in the absence of any re-acceleration processes within the lobes. The location of these areas where pressure equilibrium can be assumed is different in each type of radio galaxy. In FRI radio galaxies, the lobes expand into the medium and age as they move away from the core, resulting in a core-brightened structure in which the jet contains younger electrons and the outer edges contain an older population of electrons ( e.g., 3C 270; \citealt{Kolokythas2015_3C270}). In FRII radio galaxies, the lobes are constituted by backflow emission left behind by advancing hotspots and age towards the core, resulting in the edge-brightened structure with older electrons on the inner tips of the radio lobes (e.g., 3C 219; \citealt{Hardcastle1998FRII}). These inner tips of FRII radio galaxies and the outer tips of FRI, where an older population of electrons is found, can be used to estimate the properties of the ambient medium, assuming that they have reached pressure equilibrium with the environment. This approach has been used in other studies, e.g. \citet{Oei2023}. 

We used the SE emission on J0020-25 (Blue region on Fig. \ref{fig:AllGalaxies}, top panel), inner tip of the south radio lobe of J0019-25 (cyan region on Fig. \ref{fig:AllGalaxies}, middle panel), and the outer tip of J0019-26 (cyan region on Fig. \ref{fig:AllGalaxies}, bottom panel) to estimate the filament pressure, magnetic field strength, and total energy density around these galaxies. We selected these regions based on the aforementioned criteria, where J0020-25 and J0019-26 represent edge-brightened radio lobes, and J0019-25 represents a core-brightened radio galaxy.

PySynch \citep{hardcastle1998magnetic} was used to obtain these estimates, with parameters summarised in Table \ref{tab:synch_function_assumptions} and injection value = 2$\alpha$+1 (where $\alpha$ is the absolute spectral index). \citet{Croston2018} investigated the particle content in FRI and FRII radio galaxies. For their FRI sample, they found that a median non-radiating to radiating particles energy ratio of $U_p/U_e = 9$ (denoted as $\kappa = 9$)  is required to maintain pressure equilibrium with the surrounding environment at the lobe mid-points under equipartition. For FRI tailed radio galaxies, a population of non-radiating particles is expected in the tail regions as they entrain material from the surrounding medium, resulting in an expected ratio of $ \kappa \gg 1$ \citep{Croston2018}. A lower energy cut-off of $\gamma_{min}$ = 10 was adopted for synchrotron electrons in the tail regions; higher cut-offs (e.g. $\gamma_{min}$ = 100 or 500) would truncate this aged population and underestimate the minimum internal pressure, $P_{min}$, relative to the environment. With the actual contribution of non-radiating particles unknown, we adopt $\kappa = 10$ for the faintest tail regions, alongside $\gamma_{min}$ = 10, yielding lower limits on the internal pressure. The resulting $P_{min}$ is subject to uncertainties in the assumed source geometry, the injection spectral index, and the adopted values of $\gamma_{min}$ and $\kappa$.

To compute the injection value, the flattest spectral index value of the source is required. We extracted the flux densities within a beam-sized circular region near the core of the galaxy to capture the peak emission. This region was specifically chosen to ensure a high signal-to-noise ratio (typically S/N $>10$), which is required for reliable in-band spectral fitting \citep{Knowles2022}. While we acknowledge that the $uv$-coverage mismatch across different frequency channels can lead to a systematic steepening of the spectral index, this effect is minimised by the use of a region near the core rather than the extended emission. Because of this, we consider the derived values to be reliable estimates of the spectral slope.

The estimated spectral index values of J0020-25, J0019-25, and J0019-26 are summarised in Table \ref{table:estimated_properties}, together with the properties of the faint regions, estimated using the values and assumptions described in this section.

\begin{table}
\caption{Basic properties of all three radio galaxies in our sample.}
\label{table:galaxy_properties}
\centering
\begin{tabular}{lccccccc}
\hline\hline
Galaxy & RA  & Dec  & $z$  & $\theta$ \\
 & [h m s] & [\degr\ ' "] & &  [ deg ]   & \\
\hline
J0020$-$25 & 0 20 41.8 & $-$25 29 22.6 & 0.1273 & --  \\ 
J0019$-$25 & 0 19 21.3 & $-$25 38 01.2 & 0.1126 & 88.3  \\ 
J0019$-$26 & 0 19 17.8 & $-$26 09 56.8 & 0.1273 & 133.6 \\
\hline
\end{tabular}
\end{table}

\begin{table}
\caption{Synchrotron function assumptions.}
\label{tab:synch_function_assumptions}
\centering
\begin{tabular}{p{0.48\linewidth}l}
\hline\hline
Definition & Assumption \\
\hline
Method & Equipartition \\
$\gamma_{\mathrm{min}}$ & 10 \\
$\gamma_{\mathrm{max}}$ & $10^{5}$ \\
Spectrum & Power-law \\
Shape & Cylindrical \\
Field strength range (T) & $10^{-12} - 10^{-7}$ \\
$U_p/U_e \ (\kappa)$ & 10 \\
\hline
\end{tabular}
\end{table}

\subsection{Intrafilament density}\label{sec:Intrafilament density}
The filament's gravitational potential pulls the galaxies into the filament with a velocity ($v_{gal}$). These galaxies then experience ram pressure, which causes the jets to bend. This model of bending due to ram pressure allows us to estimate the density of the intrafilamentary medium (IFM). We adopted the approach used by \citet{edwards2010first} to estimate the density of the filament ($\rho_\text{IFM}$) by using the Euler equation as defined in \citet{begelman1979twin} and \citet{freeland2008bent}:

\begin{equation}
    \rho_\text{IFM} = \frac{\omega\Gamma^2\beta^2h}{Rv_{gal}^2}
    \label{eq: IFM density}
\end{equation}

\hfill

\noindent where $\omega$ is the enthalpy and depends on the pressure inside the jet ($\omega = 4P_{min}$), $\Gamma$ is the Lorentz factor defined by $(1-\beta^2)^{-1/2}$ where $\beta = v/c$, $h$ is the jet width, and $R$ is the radius of curvature defined by $R = b/[cos(180-a)/2]$, following the assumptions of \citet{Laing2014}, where $b$ is the length at which the jet starts to bend and $a$ is the bending angle. Here, $a$ is defined as the angle of deviation of one jet from its expected collinear path, which means $a$ = 180\textdegree\ - $\theta$ (as defined in Sect.~\ref{sec:Bending Angle}). We adopted the pressure estimates from Table \ref{table:estimated_properties} for the values of $P_{min}$. To derive the jet speed and galaxy orientation, we used the jet-counterjet ratio (denoted as $R_{\text{j}}$ throughout this work to avoid confusion with the radius of curvature) of the galaxies: 

\begin{equation}
    \beta \cos\theta = \frac{R_{\text{j}}^\delta -1}{R_{\text{j}}^\delta +1} 
\end{equation}

\hfill 

\noindent where $\delta = 1/(2+\alpha)$, and $\alpha$ is the absolute spectral index, $\theta$ is the inclination angle of the jet relative to the observer's line of sight. With the estimated values of $R_{\text{j}}$ and $\alpha$, we could estimate $\beta \cos\theta$, from which we estimated $\theta$ for different values of $\beta$, i.e $\beta$ = 1, 0.99, 0.5, 0.3, and 0.2. These estimates give the lowest and highest $\theta$ values, as well as an indication of $\beta$ estimates for our galaxies.   

The velocity of a galaxy relative to the filament, $v_{gal}$, was estimated using the line-of-sight velocity dispersion ($\sigma_v$) of galaxies within a 1 Mpc region centred around the galaxy of interest as $v_{gal}$ = $\sqrt{3}\sigma_v$. To estimate $\sigma_v$, we converted catalogued galaxy redshifts ($z_i$) into rest frame velocities ($v_{r,i}$) using the following formula \citep{Danese1980A&A},

\begin{equation}
    v_{r,i} = c \left( \frac{z_i - \bar{z}}{1 + \bar{z}} \right)
    \label{eq:Relative Velocity}
\end{equation}

\hfill 

where $\bar{z}$ is the mean redshift of galaxies within the region of interest. We scaled the corresponding measurement uncertainties as $\sigma_{v,i} = \sigma_{cz,i}/(1+\bar{z})$ in the rest frame. For the 2dFRGS sources, individual measurement errors are not catalogued; we therefore adopted the standard rms uncertainty of 85 km s$^{-1}$ for a reliable quality parameter of $Q \geq 3$ \citep{colless20012df}. We followed the work of \citet{Danese1980A&A}, where we subtract the contribution of individual measurement errors as follows, 

\begin{equation}
    \sigma_v^2  = \sigma_{v_{r,i}}^2 - \frac{1}{N} \sum_{i=1}^{N} \sigma_{v,i}^2
    \label{eq:velocity dispersion}
\end{equation}

\hfill 

where $\sigma_{v_{r, i}}$ is the sample standard deviation of the rest frame velocities ($v_{r, i}$). Although the authors followed this procedure to estimate $v_{gal}$, it is worth noting that unlike \citet{freeland2008bent} who estimated $\sigma_v$ for gravitationally bound galaxy groups, sources in this study reside in unvirialised large scale filaments, where a single high speed galaxy can contaminate the velocity estimate and the bulk motion of the filament itself can introduce systematic uncertainties, often leading to overestimating $\sigma_v$. The approach followed also assumed an isotropic velocity distribution ($\sqrt{3}\sigma_v$) while the filaments are highly anisotropic, meaning that the true velocity dispersion can deviate substantially from the estimated $v_{gal}$.

\section{Analysis and results}
\subsection{Spectral profile of J0020-25} 
\begin{figure*}
  \centering
  \includegraphics[width=0.49\textwidth]{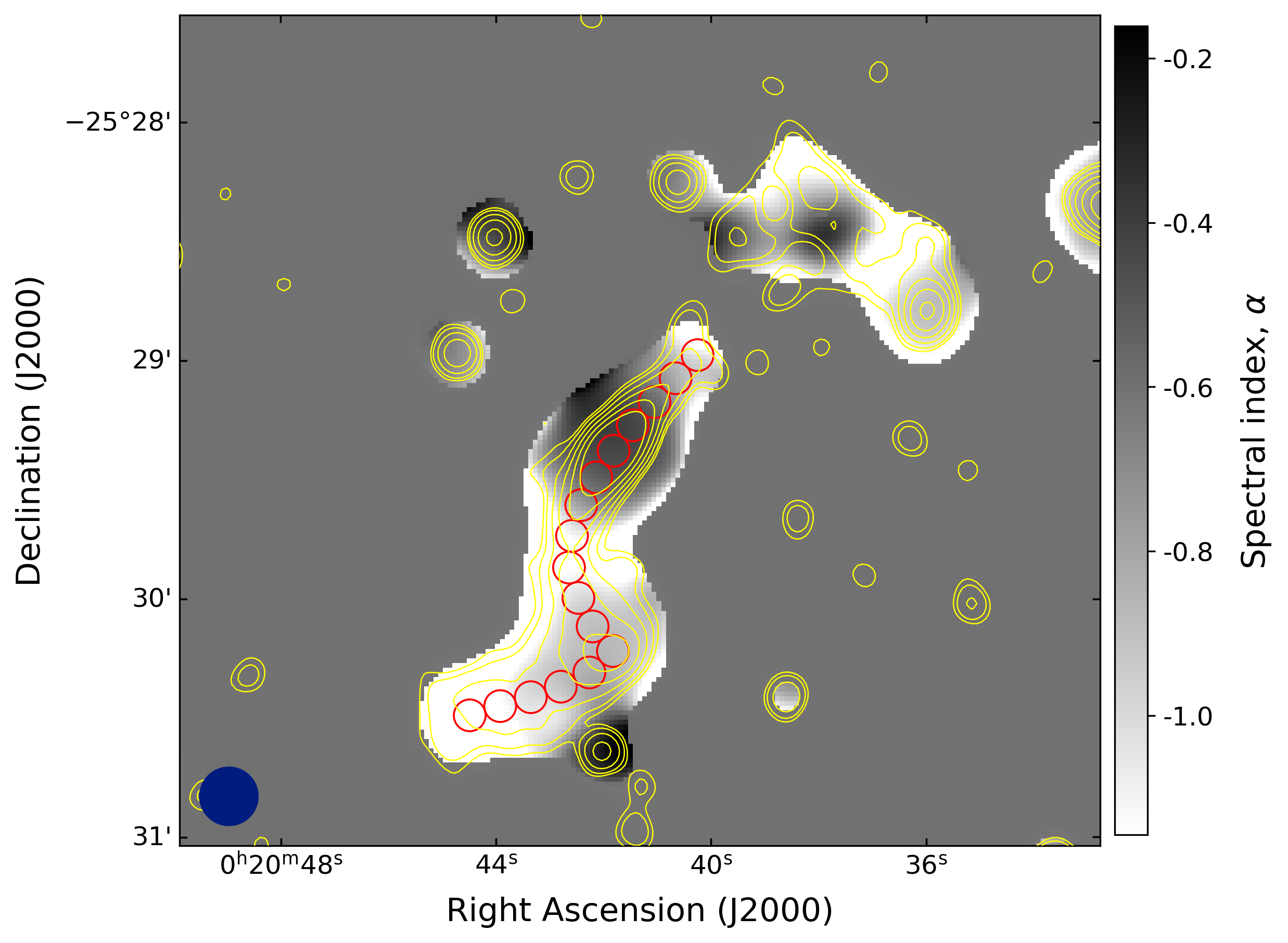}\hfill
  \includegraphics[width=0.47\textwidth]{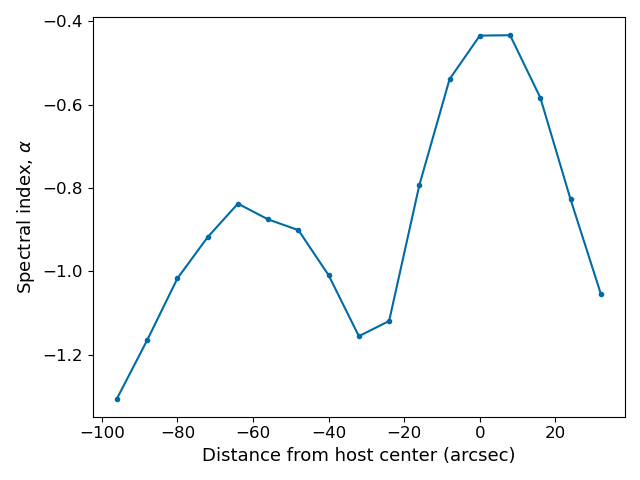}
  \caption{Left: Spectral index map of J0020-25 at the convolved resolution of 15", overlaid with full resolution radio contours at $6.6\ \mu$Jy/beam $\times$ 3, 5, 10, 20, 40, and 80 levels. Red circles indicate regions used to estimate the spectral index at different positions from the galaxy centre. Right: The spectral index variations along the galaxy, measured within the red circular regions presented on the left panel. Negative values represent the distance to the south of the host, and positive values represent the distance to the north. Zero is the position of the host centre.}
  \label{fig:Spectral index distribution}
\end{figure*}

To trace spectral index variations along J0020-25, we extracted the mean spectral index values from the convolved 15" resolution spectral index map provided in the MGCLS enhanced data products. These were measured in circular regions of 4" radius placed along the source, as indicated by the red circles in Figure \ref{fig:Spectral index distribution}. The resulting spectral index profile is shown in the right panel of this figure.

The spectral profile reveals a flat spectrum at the galaxy core ($\alpha \approx -0.43$) and a local flattening at the southern lobe hotspot. The flat spectrum at the core indicates a young, energetic electron population, while the localised flattening at the southern hotspot points to particle re-acceleration, likely driven by shock due to the jet's impact with the surrounding medium. Spectral steepening occurs between the core and the southern hotspot, consistent with energy loss as electrons travel away from the galaxy's core before re-energising at the hotspots, creating the observed radio lobes. Beyond the southern hotspot, the spectrum steepens further into the diffuse emission towards the South West, indicating an ageing electron population.

\subsection{IFM density around J0019-26}
J0019-26 was used to estimate the density of the filament along J0019-26 and J0020-25. To estimate the velocity of J0019-26 relative to the filament, we considered the velocities of galaxies within a 1 Mpc radius circular region centred on J0019-26 with $z = 0.1263 - 0.1286$. Following the procedure described in Section \ref{sec:Intrafilament density} and taking $v_{gal}$ = $\sqrt{3}\sigma_v$, we obtained a relative velocity of $v_{gal}$ = 384 $\pm$ 157 km s$^{-1}$ for J0019-26 using 4 galaxies within the specified region. The small sample size used contributes significantly to the high uncertainty of $\sim$ 41\% in the estimated value of $v_{gal}$. To estimate the jet-counterjet ratio, we used two circular regions of the same size and distance from the core, and a value of $R_{\text{j}}$ = 1.05 was found. With this value of $R_{\text{j}}$ and a spectral index of $\alpha = -0.41$ estimated using a power-law fit as explained in Section \ref{subsec:spectral analysis}, we found $\theta \in [87.2^\circ, 89.4^\circ]$, for which we have an estimated average value of $\beta$ = 0.41. 

The jet width, h, was determined by measuring the width of the 5$\sigma$ contour where the jet has a stable, well-collimated morphology, avoiding the downstream regions where the jet begins to disrupt. Using the estimated jet parameters of J0019-26 summarised in Table \ref{tab:J0019-26 estimates}, $\beta$ = 0.41, and the pressure stated on Table \ref{table:estimated_properties}, we found IFM density of $\rho_\text{IFM}$ $\approx$ 3.8$\times 10^{-29}$ g cm$^{-3}$. The estimated IFM density is about 4 times greater than the critical density of $\rho_{c}$ = 9.7 $\times 10^{-30}$ g cm$^{-3}$ at the redshift of J0019-26.

\begin{table}
\caption{Summary of J0019-26 jet parameters used to estimate the IFM density.}
    \label{tab:J0019-26 estimates}
    \centering
    \begin{tabular}{ll}
    \hline
    \hline
        Parameter  & Value \\
        \hline
         Jet width ($h$) & 50 kpc \\
        Length at which the jet starts to bend ($b$) & 37 kpc \\
         Bending angle ($a$) & 46.4\textdegree\ \\
         \hline
    \end{tabular}
\end{table}

\subsection{IFM density around J0019-25}
Following the same approaches used for J0019-26, we found $v_{gal}$ = 422 $\pm$ 105 km s$^{-1}$ using 9 galaxies found within a 1 Mpc diameter region centred on J0019-25, with the redshift limit of $z = 0.1106 - 0.1146$. We estimated a jet-counterjet ratio of $R_{\text{j}}$ = 1.72. With this value of $R_{\text{j}}$ and an estimated spectral index of $\alpha = -1.18$, we found a wide range of $\theta \in [64.9^\circ, 85.1^\circ]$ for the orientation of J0019-25, with an average value of $\beta$ = 0.40. A direct measurement of the jet width for J0019-25 was not feasible due to the unresolved nature of the jet at the current angular resolution. Consequently, we put constraints on the jet width by estimating the lower and upper limits. For the lower limit, we measured the width of the 5$\sigma$ contour of the emission bridge connecting the host AGN and the southern radio lobe. For the upper limit, we adopted the 5$\sigma$ contour width of the radio lobe. With the lower and upper limits of $h$ = 13 and 63 kpc, $b$ = 10 kpc, and $a$ = 91.7\textdegree\, we found IFM density limits of $\rho_\text{IFM}$ $\approx$ $(0.2 - 1.1)\times 10^{-26}$ g cm$^{-3}$.

\begin{table*}
\caption{Estimated properties of J0020-25, J0019-25, and J0019-26.}
\label{table:estimated_properties}
\centering
\small
\begin{tabular}{lccccccc}
\hline\hline
Galaxy & Flux & $L_{0.150\,\mathrm{GHz}}$ &  $\alpha$ & $B_{\mathrm{eq}}$  & $U_{\mathrm{eq}}$  & $P_{\mathrm{eq}}$ &$\rho_\text{IFM}$  \\
  & [mJy] & [W\,Hz$^{-1}$] &  &[$\mu$G] & [$\times 10^{-8}$\,erg\,cm$^{-3}$] & [fPa]  &[g cm$^{-3}$] \\
\hline
J0020$-$25  & 0.69 & $4.11\times 10^{24}$ &  $- 0.42 \pm$ 0.02 & 2.2 & 4.02 & 13.4 & -- \\ 
J0019$-$25 & 0.06 & $2.25\times 10^{25}$ & $-1.18 \pm$ 0.21  & 15.2 & 183.1 & 610  &$1.1\times 10^{-26}$  \\ 
J0019$-$26 & 0.56 & $8.19\times 10^{24}$ &  $-0.41 \pm$ 0.03 & 2.2 & 3.9 & 13.0  &  $3.8\times 10^{-29}$ \\
\hline
\end{tabular}
\tablefoot{The luminosities were estimated using $5\sigma$ regions, where $\sigma = 6.6\,\mu$Jy\,beam$^{-1}$. The flux density, $B_{\mathrm{eq}}$, $U_{\mathrm{eq}}$, $P_{\mathrm{eq}}$, and $\rho_\text{IFM}$, are the estimated properties of the faintest regions used to infer the IFM properties.}
\end{table*}

Table \ref{table:estimated_properties} summarises the estimated properties of J0020-25, J0019-25,
and J0019-26, which include the estimated pressure and density of the IFM. Literature works \citep{Feretti1992, Bruno2024} report that the lobes of radio galaxies in clusters are under-pressured with respect to the intracluster medium. Therefore, we may consider the values we obtained for the IFM as lower limits.

\section{Discussion}
By looking at the galaxy distribution of galaxies in the direction of the wall identified by \citet{pimbblet2006campanas} (up to $z = 0.15$), we identified and confirmed the existence of a large-scale filament at the redshift of $z \sim 0.110$ to 0.114. We further identified another filament vertically inclined in the North East direction at $0.12 < z < 0.13$, and an inclined wall joining these two filaments in the North West direction (Fig. \ref{fig:3D distribution}). Within these filaments, we identified bent-tailed radio galaxies J0020-25, J0019-26, and J0019-25.  

To date, only a few bent-tailed radio galaxies have been identified within filamentary structures. These include the discovery by \citet{edwards2010first} and the analysis of sources in the COSMOS field by \citet{Vardoulaki2021}, who further suggest that we mainly find FRI-type radio sources in filaments. Other studies like \cite{deVos2021} have identified bent-tailed radio galaxies beyond the cluster virial radii, which they claim to be within infalling filaments. The lack of discoveries of filamentary bent-tailed galaxies makes the discoveries in this work an important milestone. 

Bending of radio galaxies in filament environment can be due to ram pressure experienced by the galaxies as they have their first infall into the nearest galaxy cluster or when they are infalling into the filament's gravitational well. \citet{Rost2021} show that infalling gas and accretion shocks towards cluster outskirts along large-scale filament creates turbulence, which can bend and disrupt radio jets as the galaxies infall into the clusters. Several simulation studies have observed overdensities in the filament spines \citep{GalarragaEspinosa2021, Rost2021, GalarragaEspinosa2024}, which can elevate ram pressure effects on the galaxies, resulting in bent radio jets. \citet{Gheller2019} and \citet{GalarragaEspinosa2021} investigated the effects of the presence of dark matter halos in the filaments and found that these dark matter halos, located mainly along the filament spines, elevate the local pressures and temperatures. Radio galaxies residing in the vicinity of these halos in the filament spines can therefore experience enhanced ram pressure, bending and shaping the radio jets.

Using the faintest regions of J0019-26 and J0019-25, under the assumption that they have reached pressure equilibrium with the surrounding medium, we estimated IFM density and pressure of their respective filaments. Our IFM density estimate around J0019-26 is comparable with the $\rho_\text{IFM}$ constrain of $(1-20) \times 10^{-29}$ g cm$^{-3}$ by \citet{edwards2010first} and in agreement with simulation predictions of WHIM density of $4\times 10^{-6}$ to $1\times 10^{-4}$ cm$^{-3}$ by \citet{dave2001baryons}, while our IFM density estimate around J0019-25 is higher than the IFM density estimates by \citet{edwards2010first} and \citet{dave2001baryons}. The ambient medium density estimate around J0019-25 is in agreement with the group medium density estimates by \citet{freeland2011intergalactic}, suggesting that this galaxy might be in a group environment. 

The higher density we found around J0019-25, consistent with the environment of a galaxy group, is a natural finding when considering that galaxy groups are often embedded within the large-scale cosmic filaments \citep{Sarron2019, Tempel2014}. Filaments serve as bridges that channel galaxies and groups towards larger clusters \citep{Bonjean2018}. Within these filamentary structures, it has been shown that we can find localised regions of overdensities \citep{Tempel2014}, which represent groups before they eventually merge into a cluster at the nodes of the cosmic web. Our results for J0019-25 likely reflect this transition, placing the galaxy in a denser group environment rather than the more diffuse IFM.  

We compared the IFM pressure estimates found for the filament medium surrounding J0020-25 (estimated using J0019-25 faintest region), J0019-25, and J0019-26 to the X-ray filaments' pressure values summarised by \citet{malarecki2013giant}, who compiled a table of studies that have estimated the electron density of X-ray emission detections from hot gas in filaments connecting galaxy clusters. We have reproduced their table in Table \ref{table:X-ray_observations}. We also compared the IFM pressure estimates to the filament pressure estimates of: (i) \citet{Subrahmanyan2008, Safouris2009}; and \citet{oei2022discovery}, who estimated intergalactic medium (IGM) pressures using galaxies associated with filamentary structures, and (ii) \citet{GalarragaEspinosa2021} who studied the properties of gas around filaments at $z = 0$ and found average pressure estimates of $4-12 \times 10^{-7}$ keV cm$^{-3}$ at the filament cores, which is roughly three orders of magnitude lower than the intracluster medium \citep{Arnaud2010}.

\begin{table}
\caption{WHIM pressure estimates from X-ray observations, radio galaxies, and numerical simulations.}
\label{table:X-ray_observations}
\centering
\begin{tabular}{lc}
\hline\hline
Reference & Pressure [$10^{-15}$ Pa] \\
\hline
Briel \& Henry (1995) &  $<8.99$ \\
Werner et al. (2008) & $5.41 \pm 3.55$  \\
Subrahmanyan et al. (2008) & $>$ 4.1\\
Safouris et al. (2009) & > 1.0 \\
Fraser-McKelvie et al. (2011) & $57.10 \pm 2.43$ \\
Dietrich et al. (2012) & $5.55 \pm 3.65$ \\
Galárraga-Espinosa et al. (2021) & 0.64 - 1.92\\
Oei et al. (2022) & 1.5 \\
\hline
\end{tabular}
\tablefoot{Radio estimates are adopted from \citet{Subrahmanyan2008, Safouris2009}; and \citet{oei2022discovery}. Simulation estimate is from \citet{GalarragaEspinosa2021}. The remaining values are X-ray estimates adopted from Table 4 of \citet{malarecki2013giant}. All pressure estimates in the table have been converted to Pascals (Pa).}
\end{table}

Filament gas pressure estimates from J0020-25 and J0019-26 lie very close and within the range of the WHIM pressure estimates summarised in Table \ref{table:X-ray_observations} from X-ray, radio, and numerical simulation studies. The pressure estimates of J0020-25 and J0019-26 are similar to one another, further supporting our assumption that these two galaxies are found within the same filamentary structure, with similar environmental conditions.  However, different morphological effects are seen on each one. We suggest this is due to the difference in their orientation relative to the filamentary environment. 

J0020-25 appears to be on the periphery of the filament. We found a jet-counterjet ratio of R = 1.95 for J0020-25, where the counterjet is the disrupted north jet. Using R = 1.95 for this galaxy, we found an inclination angle constraint of $\theta \in$ [46.6, 82.1]. For this inclination angle constraint, we have an average value of $\beta$ = 0.39, which corresponds to an inclination angle of $\theta$ = 69.5\textdegree. This inclination angle suggests that the system lies nearly parallel to the plane of the sky, minimising deprojection uncertainties and allowing for a more confident mapping of the jet's trajectory. Given the location of this galaxy on the periphery of the filament, the disruption of the northern lobe could be attributed to ram pressure or interaction with the local IGM as the galaxy moves through this denser medium. In this scenario, the filament provides a source of external pressure that could disrupt the receding jet.

J0019-25, located in a different filamentary structure at the redshift of $z = 0.1126$, has an estimated filament gas pressure higher than the X-ray WHIM pressure estimates (Table \ref{table:X-ray_observations}). High-pressure and density estimates inferred for the surrounding medium of this galaxy may account for the observed high bending effects, indicating that it is experiencing strong ram pressure effects. A greater degree of bending in radio galaxies is generally associated with higher-density regions or high ram pressure. Based on pressure and density estimates of the surrounding medium, we infer that J0019-25 might reside within a group environment. While the diffuse WHIM within filaments may induce subtle jet curvature via bulk flow, the transition into the dense intragroup medium results in significant ram pressure stripping of the galaxy’s hot gas and a characteristic sharpening of jet bending. This allows the greater bending effects to serve as sensitive probes of the transition between large-scale filaments and virialised groups. 

\begin{figure}
    \centering
    \includegraphics[width=0.99\linewidth]{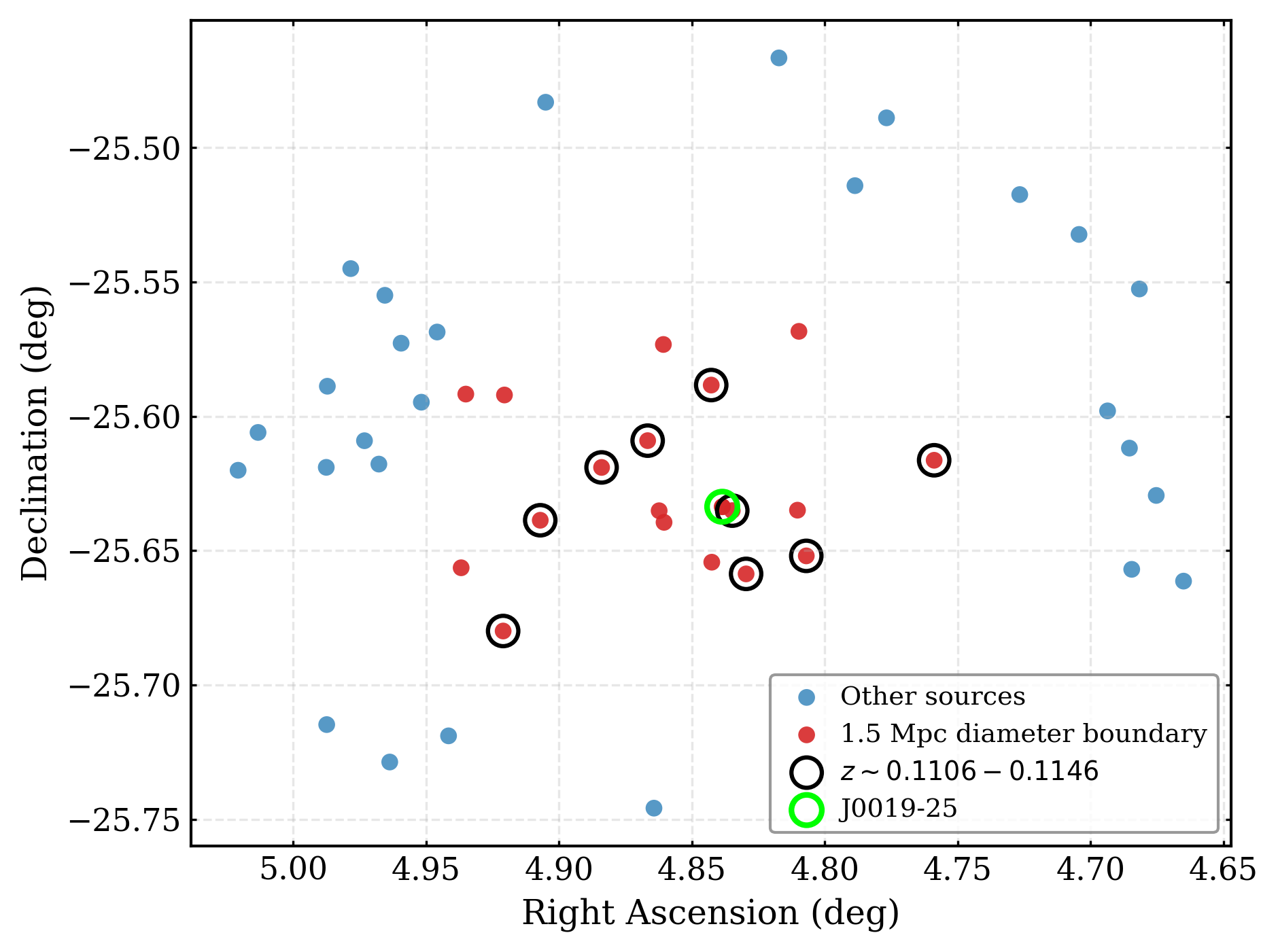}
    \caption{2D distribution of galaxies around J0019-25. Red circles indicate spectroscopic members within a projected diameter of 1.5 Mpc ($\approx$ 0.2$\degr$) centred on the brightest group galaxy (BGG) candidate J0019-25 (highlighted by a green circle). Potential group members within 0.1106 $\leq z \leq$ 0.1146 are indicated by black circles, suggesting a coherent group structure.}
    \label{fig:J0019-25 neighbours}
\end{figure}

We further looked into the brightness of neighbouring galaxies (Fig. \ref{fig:J0019-25 neighbours}), where we discovered that J0019-25 is the brightest galaxy. This identifies J0019-25 as a potential Brightest Group Galaxy (BGG), observed well after its initial transition into the group environment. To identify potential group members, we applied the small-angle approximation ($\theta \approx d/D_A$) to constrain the RA and Dec for sources within 1.5 Mpc projected diameter of J0019-25 (Fig. \ref{fig:J0019-25 neighbours}). By restricting the redshift range to $z = z_{\text{J0019-25}} \pm 0.006$, we recovered 11 candidate members. We converted individual galaxy redshifts ($z_i$) into line-of-sight velocity offsets relative to the potential BGG (J0019-25) using Equation \ref{eq:Relative Velocity}, where $\bar{z} = z_{\text{BGG}}$. The rest frame line-of-sight velocity dispersion of our potential group ($\sigma_{v,g}$) was then estimated using Equation \ref{eq:velocity dispersion}.

Applying a standard 3$\sigma$ cut excludes one outlier at $z = 0.1091$ from the 11 potential group members, yielding a line-of-sight rest frame velocity dispersion of $\sigma_{v,g} =255\pm 60$ km.s$^{-1}$ (N=10) versus $\sigma_{v,g} =352\pm 78$ km.s$^{-1}$ (N=11). This velocity dispersion is consistent with that of a typical galaxy group \citep{Zabludoff_1998, Hou2013}. Assuming an isotropic gas, this yields a group dispersion velocity of $v_{v,g} = \sqrt{3}\sigma_{v,g} =441\pm 104$ km.s$^{-1}$. However, the small number of members introduces uncertainty into this dispersion estimate. The final sample of N = 10 potential group members contains galaxies within $z_{\text{J0019-25}}$ $\pm$ 0.002 (0.1106 $\leq z \leq$ 0.1146), as indicated by black circles in Figure \ref{fig:J0019-25 neighbours}. 

\section{Implications and conclusions}
Our work confirmed the existence of the `wall' identified by \citet{Pimbblet2005} at the redshift of $z\in[0.111, 0.115]$. We identified a bent-tailed radio galaxy within this filament and two more bent-tailed radio galaxies within a separate filamentary structure at the redshift of $z\sim 0.1273$. These discoveries show the capabilities of using bent-tailed radio galaxies as probes to locate the otherwise invisible large-scale filaments. Together with optical data, these radio detections allow us to probe the properties of the Warm-Hot Intergalactic Medium (WHIM), which makes the targeted radio follow-up of known optical filamentary structures a compelling prospect. 

Beyond serving as probes for the WHIM, these bent-tailed radio sources provide a unique radio-continuum perspective on the preprocessing of galaxies. Their distorted shapes confirm that the filamentary environment is active enough to shape galaxy evolution during the transition from the field to the cluster, reinforcing the importance of the cosmic web in the life cycles of member galaxies.

To our knowledge, this is the second detection of radio galaxies interacting with a large-scale filament to date (the first and confirmed detection was found by \citealt{edwards2010first}). These sources are currently rare and highlight the need to move away from the paradigm that bent-tailed galaxies, especially WATs, could only correspond with clusters of galaxies. The rarity of these sources is no surprise due to two reasons: 

(i) In low density filament environment, the galaxy should be moving at very high speeds to experience jet bending due to ram pressure.

(ii) It is difficult to detect the faint diffuse filamentary gas (i.e. WHIM) and identify true large-scale filamentary structures with current instruments.

As the sensitivity of radio and ancillary observations improves, we predict that more such sources in filaments will become known and may serve as valuable probes of WHIM media, with which filaments are associated. In particular, we note that one of the galaxies shows what seems to be a disrupted lobe, and we connect this to possible asymmetries in the filamentary environment, indicating a denser intrafilamentary medium to the north of the galaxy. One possible interpretation of this is that the host galaxy lies near the periphery of the filament, such that the northern lobe expands into the denser filament medium while the southern lobe expands towards a less dense field environment. For BTRGs without asymmetry, the jet bending angles provide the most reliable probes of critical physical parameters of the intergalactic medium in groups and filaments. 

We further presented a detailed analysis of three bent-tailed galaxies in a filamentary environment, highlighting the interplay between their morphologies and the properties of the surrounding medium. We found IFM density of $\rho_{IFM}$ $\approx 3.8 \times 10^{-29}$ g cm$^{-3}$ for J0020-25 and J0019-26 surrounding medium and $\rho_{IFM}$ $\approx (0.2 - 1.1) \times 10^{-26}$ g cm$^{-3}$ around J0019-25. We found that our $\rho_{IFM}$ estimates for the intrafilament environment around J0020-25 and J0019-26 lie within the literature estimates, while the density estimate of the environment around J0019-25 is higher than the literature IFM density estimates and in agreement with the density of a group environment. The local density and brightness of J0019-25 are consistent with those of a BGG, suggesting that J0019-25 resides in a galaxy group embedded within the filament.

To confirm the nature of the identified filamentary BTRGs, their interaction with the large-scale structure, and hence their use as reliable probes of the ambient medium, several multi-wavelength studies will be required:  

(i) VLBI observations will resolve the compact cores, jets, and hotspots of J0020-25 and J0019-25, refining their current morphological classifications as a disrupted-lobe system and a WAT FRII radio galaxy.

(ii) Targeted X-ray observations/stacking can provide direct measurements of the thermal plasma. Such detections are more likely for J0019-25, which is located within a dense galaxy group. These measurements will constrain the ambient electron density, the thermal pressure, and the ambient mass density, which -- when combined with radio non-thermal pressure estimates -- can be used to constrain the host galaxy's velocity.

(iii) Stacking of the thermal Sunyaev-Zel’dovich (tSZ) effect will confirm the existence of the identified filaments and measure the thermal gas pressure. Similar to X-ray observations, this pressure can be used to estimate the gas density and the galaxy’s velocity through the filament or group. Moreover, identifying the filament spine will precisely pinpoint the spatial positions of the BTRGs within the broader filament structure.

(iv) Polarisation follow-up studies will build upon the weak polarisation currently recovered in the MGCLS products, providing critical insights into the magnetic field properties of both the BTRGs and the surrounding medium to further refine our understanding of their complex interactions.

We emphasise that using radio galaxies to estimate the properties of the ambient medium relies on our ability to detect faint diffuse emission in the radio lobes and jets. This requires both high sensitivity and good angular resolution. Future instruments like the SKA, especially the SKA-Low, are expected to provide the necessary sensitivity and resolution to detect such faint diffuse emission. 

Our work has focused on BTRGs located in the Southern hemisphere, which suffers from relatively poor ancillary data coverage compared to the Northern hemisphere. This situation is set to improve with the advent of instruments like the Vera Rubin Telescope, which will provide much-needed optical and infrared data complementary to the observations of telescopes like MeerKAT, ASKAP and eventually the SKA. 

\section*{Data availability}
The data used for this work, MGCLS data products release (DR1), is publicly available here\footnote{\url{http://mgcls.sarao.ac.za/data-releases/}}. \\

\begin{acknowledgements}
The authors thank the anonymous referee for very helpful comments, which helped shape the paper for the better. The financial assistance of the South African Radio Astronomy Observatory (SARAO) towards this research is hereby acknowledged (\href{https://www.sarao.ac.za}{www.sarao.ac.za}). The MeerKAT telescope is operated by SARAO, which is a facility of the National Research Foundation (NRF), an agency of the Department of Science and Innovation. K Kolokythas acknowledges the funding support from SARAO and the NRF (grant UID: 97930). \\

We also acknowledge the use of the ilifu cloud computing facility – www.ilifu.ac.za, a partnership between the University of Cape Town, the University of the Western Cape, Stellenbosch University, Sol Plaatje University and the Cape Peninsula University of Technology. The ilifu facility is supported by contributions from the Inter-University Institute for Data Intensive Astronomy (IDIA – a partnership between the University of Cape Town, the University of Pretoria, and the University of the Western Cape), the Computational Biology division at UCT, and the Data Intensive
Research Initiative of South Africa (DIRISA). \\

This work made use of the CARTA (Cube Analysis and Rendering Tool for Astronomy) software (DOI \href{https://zenodo.org/records/18477253}{10.5281/zenodo.3377984} - \url{https://cartavis.github.io}).
\end{acknowledgements}

\bibliographystyle{aa} 
\bibliography{example}

\end{document}